\documentclass[twocolumn,pra,aps,nofootinbib,superscriptaddress]{revtex4-1}
\usepackage{amsfonts}
\usepackage{amssymb}
\usepackage{mathrsfs}
\usepackage{amsthm}
\usepackage{graphicx}
\usepackage{amsmath}
\usepackage{color}
\usepackage{tikz}
\newcommand{\cc}{\mathbb{C}}

\newcommand{\hh}{\mathcal{H}}

\newcommand{\psii}[1]{\ket{\psi_{#1}}}
\newcommand{\fif}[1]{\ket{\varphi_{#1}}}

\newcommand{\1}{\ket{1}}
\newcommand{\2}{\ket{2}}
\newcommand{\3}{\ket{3}}

\newtheorem{Proposition}{Proposition}
\newtheorem{thm}{Theorem}

\theoremstyle{definition}

\newcommand{\Tr}[0]{\mathrm{Tr}}
\newcommand{\ot}[0]{\otimes}
\newcommand{\bei}{\begin{itemize}}
\newcommand{\eei}{\end{itemize}}
\newcommand{\ket}[1]{|#1\rangle}

\def\<{\langle}
\def\>{\rangle}
\newcommand{{\Cn}}{{\mathbb{C}^4}}
\newcommand{{\CN}}{{\mathbb{C}^{2n}}}
\newcommand{{\BC}}{{\mathcal{B}(\mathbb{C}^n)}}
\newcommand{{\BBC}}{{\mathcal{B}(\mathbb{C}^{2n})}}
\def\oper{{\mathchoice{\rm 1\mskip-4mu l}{\rm 1\mskip-4mu l}{\rm 1\mskip-4.5mu l}{\rm 1\mskip-5mu l}}}

\begin{document}

\title{A class of Bell diagonal entanglement witnesses in $\mathbb{C}^4 \otimes \mathbb{C}^4$: \\ optimization and the spanning property}


\author{Anindita Bera}
\affiliation{Institute of Physics, Faculty of Physics, Astronomy and Informatics,
Nicolaus Copernicus University, Grudzi\c{a}dzka 5/7, 87--100 Toru{\'n}, Poland}
\author{Filip A. Wudarski}
\affiliation{Quantum Artificial Intelligence Lab. (QuAIL), Exploration
Technology Directorate, NASA Ames Research Center,
Moffett Field, CA 94035, USA}
\affiliation{USRA Research Institute for Advanced Computer Science
(RIACS), Mountain View, CA 94043, USA}
\author{Gniewomir Sarbicki}
\affiliation{Institute of Physics, Faculty of Physics, Astronomy and Informatics,
Nicolaus Copernicus University, Grudzi\c{a}dzka 5/7, 87--100 Toru{\'n}, Poland}
\author{Dariusz Chru{\'s}ci{\'n}ski}
\affiliation{Institute of Physics, Faculty of Physics, Astronomy and Informatics,
Nicolaus Copernicus University, Grudzi\c{a}dzka 5/7, 87--100 Toru{\'n}, Poland}

\begin{abstract}
Two classes of Bell diagonal indecomposable entanglement witnesses in $\mathbb{C}^4 \ot \mathbb{C}^4$ are considered. 
Within the first class, we find a generalization of the well-known Choi witness from $\mathbb{C}^3 \ot \mathbb{C}^3$, while the second one contains the reduction map.
Interestingly, contrary to  $\mathbb{C}^3 \ot \mathbb{C}^3$ case, the generalized Choi witnesses are no longer optimal. We perform an optimization procedure of finding spanning vectors, that eventually gives rise to optimal witnesses. 
 Operators from the second class turn out to be optimal, however, 
without the spanning property. This analysis sheds a new light into the intricate structure of optimal entanglement witnesses.

\end{abstract}

\maketitle

\section{Introduction}
Quantum entanglement is arguably the most peculiar feature of quantum theory that also demarcates it from the classical one \cite{NC,HHHH,das1}. It is a crucial ingredient in quantum information processing applications, such as (entanglement-based) quantum cryptography~\cite{cripto1}, quantum dense coding~\cite{dc1}, quantum
teleportation~\cite{tel1}, measurement-based computation~\cite{briegel1} etc. However, despite tremendous progress in understanding of entanglement, we still lack its full characterization, especially in terms of entanglement detection, i.e. methods that can faithfully discriminate between entangled and separable states. 
Indeed, a wealth of various operational entanglement criteria for bipartite quantum states already exist in the literature, in particular, the partial transposition criterion \cite{Peres,HHH1}, the majorization criterion~\cite{m1}, the cross-norm or realignment criterion~\cite{cr1,cr2,cr3}, the covariance matrix criterion~\cite{cm1,cm2} to name a few.  
However, due to a very rich structure of entanglement, most of currently available techniques are either only sufficient in characterization or restricted to a narrow class of states \cite{NC,HHHH,das1}. Therefore, it is of paramount importance to have robust ways of entanglement detection, in particular operational (i.e. measurable) tools are of special interest. One such method is based on the notion of entanglement witness (EW) - an observable that is capable of detecting entangled states \cite{HHH1,HHH2,Terhal1,Terhal2}. 
The central concept of EWs is based on the ideas of the Hahn-Banach theorem on normed linear spaces~\cite{norm1}. 
A special subclass of EWs called optimal entanglement witnesses (OEWs) is considered {\it the gold standard} for measurable entanglement detection, since the observables that are OEWs allow to detect the largest set (in terms of its cardinality) of entangled states \cite{Lew,Lew2,Sperling}. Thus, making them a suitable choice for investigation.  

In this paper, we analyze a family of EWs  in $\mathbb{C}^4 \otimes \mathbb{C}^4$ which are Bell diagonal and covariant w.r.t. maximal commutative subgroup of the unitary group $U(4)$. This is a highly symmetric family that can be analyzed in detail. 
We consider a  class of EWs in $\Cn \ot \Cn$ displaying a characteristic circular structure. Such class of circulant witnesses in $\mathbb{C}^3 \otimes \mathbb{C}^3$ without any connection to entanglement theory was proposed in \cite{Cho-Kye} as a generalization of a seminal indecomposable positive map of Choi \cite{choi1}. Further analysis was provided in \cite{FilipI,Kye-PRA,Gniewko}. In particular, it was shown that these witnesses are optimal \cite{Kye-PRA,Gniewko}. In our paper, we show that the corresponding class of circulant witness  $\Cn \ot \Cn$ splits into two subclasses:  one class provides a generalization of the Choi witnesses in $\mathbb{C}^3 \ot \mathbb{C}^3$ and the other contains the witness corresponding to the reduction map. Interestingly, contrary to  $\mathbb{C}^3 \ot \mathbb{C}^3$ the generalized Choi witnesses are no longer optimal. Following a general framework \cite{Lew,Lew2} (see also \cite{Sperling} for a slightly different approach) we perform an optimization procedure which eventually gives rise to optimal witnesses satisfying the spanning property. Due to the symmetry of the considered class, the optimization procedure can be performed analytically. We show that the witnesses from the second class are optimal, however, without the spanning property. We believe that our analysis in $\mathbb{C}^4 \ot \mathbb{C}^4$ sheds a new light into the structure of optimal entanglement witnesses and can find applications in experimental entanglement detection.

This paper is organized as follows. In Sec.~\ref{prel2}, we briefly introduce  the theory of the optimal entanglement witness. In Sec.~\ref{cov3}, we present covariant Bell diagonal entanglement witnesses in $\mathbb{C}^n \otimes \mathbb{C}^n$, and also describe the corresponding  entanglement witness in $3 \otimes 3$ and $4 \otimes 4$. The main results on optimality of the introduced classes is given in Sec.~\ref{optimal4}. Finally, in  Sec.~\ref{conclude} we provide concluding remarks. 

\section{Preliminaries}
\label{prel2}
The most general approach to discriminate between separable and entangled states of a quantum composite system living in $\mathcal{H}_A \ot \mathcal{H}_B$ is  based on the notion of positive {(but not completely positive)} maps or equivalently
 entanglement witnesses (EWs) \cite{HHHH,das1,Terhal1,Terhal2,HHH1,HHH2,Guhne,TOPICAL,ani}.  A state represented by a density operator $\rho$ in $\mathcal{H}_A \ot \mathcal{H}_B$ is separable iff it can be represented as a mixture of product states $\rho = \sum_k p_k \rho_A^{(k)} \otimes \rho_B^{(k)}$, with $\rho_A^{(k)}$ and $\rho_B^{(k)}$ being density operators of subsystems A and B, respectively {\cite{werner}}.
Recall, that a Hermitian operator $W$ acting on $\mathcal{H}_A \ot \mathcal{H}_B$ is an entanglement witness (EW) \cite{Terhal1,Terhal2} if $\< \psi \ot \phi|W|\psi \ot \phi\> \geq 0$ but $W$ is not a positive operator (cf. also \cite{ultra,mirror} for the concept of the ultra EW and mirror EW, respectively). A state $\rho$ is entangled if and only if there exists an EW $W$ such that ${\rm Tr}(W \rho)<0$. 
Equivalently, a bipartite state $\rho$ living in $\hh_A\otimes\hh_B$ is entangled iff there exists a positive map $\Phi$ such that $(\oper_A \ot \Phi)\rho$ is no longer a positive operator ($\oper_A$ denotes identity map acting on subsystem A). 

In qubit-qubit or qubit-qutrit case the situation is fully solved: a state $\rho$ is separable if and only if it is PPT, i.e. its partial transposition is positive ($\rho^\Gamma =(\oper_A \otimes T)\rho \geq 0$) \cite{Pawel,Peres}. However, in general the so-called separability problem is notoriously hard, that is, there exist PPT states which are entangled (they belong to the class of so-called bound entangled states \cite{HHHH}).

Recall, that an EW $W$ is decomposable if 
\begin{equation}\label{eq:decomposable}
    W = A + B^\Gamma,
\end{equation}
where $A,B \geq 0$ and $\Gamma$ denotes the partial transposition.  It is clear from the definition that a decomposable EW is unable to detect an entangled PPT state. Therefore, it is more interesting to examine EWs that are of indecomposable form, i.e. they fail to be represented in the form of  Eq.~\eqref{eq:decomposable}. 
Equivalently, a linear positive map $\Phi$ is decomposable if $\Phi = \Lambda_1 + \Lambda_2 \circ T$, where $\Lambda_1,\Lambda_2$ are completely positive and $T$ denotes transposition.

Given an EW, one may define a set of entangled states ${\cal D}_W$ detected by $W$, that is, ${\cal D}_W =\{ \rho\ |\ {\rm Tr}(W \rho) <0\}$. It is {straightforward}  that if ${\cal D}_W \supset {\cal D}_{\tilde{W}}$, then $W$ is more effective than $\tilde{W}$, since it detects more entangled states than $\tilde{W}$ does. Following the Refs.~\cite{Lew,Lew2}, an entanglement witness $W$ is optimal if there is no other EW $W'$ such that ${\cal D}_W \subset {\cal D}_{W'}$. 
Therefore the knowledge of optimal EWs is sufficient for the full characterization of separable/entangled states.

In this regard, the following sufficient condition for optimality was provided in \cite{Lew}: if a set of product vectors $|\psi_k \otimes \phi_k\rangle$ satisfying

\begin{equation}\label{!}
 \< \psi_k \ot \phi_k|W|\psi_k \ot \phi_k\> = 0 ,
\end{equation}
span $\hh_A \otimes \hh_B$, then $W$ is optimal. And therefore, one can say that such $W$ has the spanning property. 
Several examples of EWs satisfying (\ref{!}) (hence optimal) {already exist} in the literature \cite{Hall,Breuer,Justyna1,Justyna11,J2,J3,Kye-PRA} (cf. also review papers \cite{Kye-Rev,Hansen}). {Since the condition \eqref{!} is only sufficient, one may identify optimal EWs without the spanning property. In \cite{Remik}, authors provided such an example for a decomposable case.}

\section{Covariant Bell diagonal entanglement witnesses in $\mathbb{C}^d \otimes \mathbb{C}^d$}
\label{cov3}

Let us consider an $n$-dimensional Hilbert space. By fixing a computational basis $\{|0\>,\ldots,|n-1\>\}$, we introduce a family of Weyl unitary operators $U_{mk}$ defined via \cite{Bell1,Bell2,Bell3}

\begin{equation}\label{}
  U_{mk} |\ell\> = \omega^{m\ell} |\ell+k\>~(\mbox{mod}~n) ,
\end{equation}
with $\omega = e^{2 \pi i/n}$. Weyl operators (matrices) satisfy

\begin{equation}\label{}
  U_{k\ell} U_{rs} = \omega^{ks} U_{k+r,\ell + s} ,
\end{equation}
together with $U_{k\ell}^* = U_{-k\ell}$, $U^\dagger_{k\ell} = \omega^{k\ell} U_{-k,-\ell}$, and ${\rm Tr}( U_{k\ell} U_{rs}^\dagger) = n\, \delta_{kr}\delta_{\ell s}$ (with all index summation being modulo $n$). One can define the generalized Bell states in $\mathbb{C}^n \otimes \mathbb{C}^n$ via

\begin{equation}\label{}
  |\psi_{k\ell}\> = \oper_d \otimes U_{k\ell} |\psi^+_n\> ,
\end{equation}
where $ |\psi^+_n\> = 1/\sqrt{d} \sum_{k=0}^{n-1} |k \otimes k\>$ stands for the canonical maximally entangled states. A bipartite operator $X$ in $\mathbb{C}^n \otimes \mathbb{C}^n$ is Bell diagonal if 
\begin{equation}\label{}
  X = \sum_{k,\ell=0}^{n-1} x_{k\ell} P_{k\ell} ,
\end{equation}
where $P_{k\ell} =|\psi_{k\ell}\>\<\psi_{k\ell}|$. Consider now a maximal commutative subgroup of $U(n)$
\begin{equation}\label{}
  T(n) = \{ \ U \in U(n)\, |\, U = \sum_{k=0}^{n-1} e^{i \phi_k} |k\>\<k| \ \} ,
\end{equation}
with $\phi_k \in \mathbb{R}$. Moreover,  a bipartite operator $X$ is said to be $T \otimes T^*$-covariant whenever
\begin{equation}\label{}
  U \otimes U^* X (U \otimes U^*)^\dagger = X ,
\end{equation}
for any $U \in T(n)$. Actually, any covariant operator has the following structure 
\begin{equation}
\label{XAB}
  X = \sum_{k,\ell=0}^{n-1} A_{kl} |k\>\<k| \otimes |\ell\>\<\ell| + \sum_{k\neq \ell=0}^{n-1} B_{kl} |k\>\<\ell| \otimes |k\>\<\ell|  ,
\end{equation}
with complex parameters $A_{k\ell}$ and $B_{k\ell}$. In this paper we are going to analyze the Hermitian Bell diagonal operators which  are $T \otimes T^*$-covariant. It turns out \cite{Bell2,Bell3,JPA} that for such operators the Hermitian matrix $A_{k\ell}$ is circulant, i.e. $A_{k\ell} = \alpha_{k-\ell}$ for some real vector $(\alpha_0,\alpha_1,\ldots,\alpha_{n-1})$. Moreover, all $B_{kl}$ are constant, i.e. $B_{k\ell} = \beta \in \mathbb{R}$.

Now, the following Proposition provides the sufficient condition for the circulant matrix $A_{k\ell}$ and the parameter $\beta$ that guarantee that (\ref{XAB}) defines a legitimate entanglement witness.


\begin{Proposition}\cite{OSID} If  the circulant matrix $A_{k\ell}$ satisfies the following constraints

\begin{equation}\label{I0}
  \alpha_0 + \alpha_1 + \ldots + \alpha_{n-1} = n-1 ,
\end{equation}
together with

\begin{equation}\label{IJ}
  AA^T = \mathbb{I} + (n-2) \mathbb{J},
\end{equation}
where $\mathbb{J}_{k\ell}= 1$,  then

\begin{equation}\label{W}
  W = \sum_{k,\ell=0}^{n-1} \alpha_{k-l} |k\>\<k| \otimes |\ell\>\<\ell| - \sum_{k\neq \ell=0}^{n-1} |k\>\<\ell| \otimes |k\>\<\ell|,
\end{equation}
defines an EW.  
\end{Proposition}
Introducing the following projectors
\begin{equation}\label{}
  \Pi_k = P_{k0} + P_{k1} + \ldots + P_{k,n-1} ,  
\end{equation}
for $k=0,1,\ldots,n-1$ formula (\ref{W}) can be rewritten in the  compact form as
\begin{equation}\label{W!}
  W = (\alpha_0+1) \Pi_0 + \alpha_1 \Pi_1 + \ldots + \alpha_{n-1} \Pi_{n-1} - n P_n^+ ,
\end{equation}
where $P_n^+ = |\psi_n^+\>\<\psi_n^+|$ stands for the projector onto the maximally entangled states. 

\subsection{EWs in $3 \otimes 3$ }
For $n=3$ let us use the following notation $a=\alpha_0,\ b=\alpha_1,\ c=\alpha_2$ and hence the circulant  matrix $A_{k\ell}$ has the structure
\begin{equation}\label{}
   A= \left(  \begin{array}{ccc} {a} & {b} & {c}  \\ {c} & {a} & {b}  \\ {b} & {c} & {a}  \end{array} \right) .
\end{equation}
Conditions  (\ref{I0}) and (\ref{IJ}) imply
\begin{eqnarray}
  a+b+c &=& 2 \\
  a^2+b^2 + c^2 &=& 2 \\
  ab+bc+ca &=& 1 , 
\end{eqnarray}
which after simple algebra gives rise to
\begin{eqnarray}
  a+b+c =2 , \ \  a^2+b^2+c^2 = 2 ,
\end{eqnarray}
or equivalently \cite{Cho-Kye}
\begin{eqnarray} \label{elipsa}
  a+b+c =2 , \ \  bc=(a-1)^2  ,
\end{eqnarray}
The corresponding EW (\ref{W!})  reads as follows
\begin{equation}  \label{W3}
W[a,b,c] = (a+1)\Pi_0 + b\, \Pi_1   + c\, \Pi_2    - P^+_3 .
\end{equation}
Hence the above class may be parameterized by a single parameter $\phi\in [0,2\pi)$ \cite{Kossak,FilipI}
\begin{eqnarray}
  a &=& \frac 23 ( 1 + \cos \phi) , \nonumber \\
   b &=& \frac 13 ( 2 - \cos \phi - \sqrt{3} \sin\phi) , \\
    c &=& \frac 13 ( 2 - \cos \phi + \sqrt{3} \sin\phi) . \nonumber
\end{eqnarray}
One proves \cite{Kye-PRA} (see also \cite{Gniewko} for another proof) that if $a \leq 1$, then  $W[a,b,c]$ defines an optimal EW.
Moreover,  if $a < 1$, then $W[a,b,c]$ enjoys spanning property \cite{Kye-PRA,Gniewko}. However, for $a=1$,  two EWs correspond to Choi maps $W[1,1,0]$ and $W[1,0,1]$ for which there are only seven vectors $|\psi_k \otimes \phi_k\rangle$ satisfying $\< \psi_k \otimes \phi_k|W[a,b,c]|\psi_k \otimes \phi_k\>=0$.

\subsection{ EWs in $4 \otimes 4$ }
For $n=4$, one has the corresponding circulant matrix
\begin{equation}\label{}
   A = \left(  \begin{array}{cccc} {a} & {b} & {c} & {d} \\ {d} & {a} & {b} & {c}  \\ {c} & {d} & {a} & {b} \\  {b} & {c} & {d} & {a}  \end{array} \right),
\end{equation}
where $d = \alpha_3$. Conditions  (\ref{I0}) and (\ref{IJ}) imply
\begin{eqnarray}  
  a+b+c+d &=& 3 \label{II-1}\\
  a^2+b^2+c^2+d^2 &=& 3 \label{II-2} \\
  ac+bd &=& 1 , \\
  (a+c)(b+d) &=& 2 .
\end{eqnarray}
Simple algebra provides two solutions: class I is characterized by (\ref{II-1})-(\ref{II-2}) together with
\begin{equation}\label{I}
  a+c=2 \ , \ \ b+d=1 \ ,
\end{equation}
whereas class II 
is characterized by (\ref{II-1})-(\ref{II-2}) together with
\begin{equation}\label{II}
  a+c=1 \ , \ \ b+d=2 \ .
\end{equation} 
The corresponding EW has the following form
\begin{equation}  \label{W4}
W[a,b,c,d] = (a+1)\Pi_0 + b\, \Pi_1   + c\, \Pi_2 + d\, \Pi_3    - P^+_4 .
\end{equation}
In the next section, we provide the detailed analysis of these two classes of EWs.

\section{Optimality}
\label{optimal4}
The key question we address in this section is whether or not EWs constructed in the previous Section for $n=4$ are optimal.   

\subsection{Class I}
For the class defined by (\ref{I}), we introduce the following parametrization
\begin{eqnarray}
a &=& \frac{1}{2} (2-\sin\theta), \nonumber\\
b &=& \frac{1}{2} (1+\cos\theta), \nonumber\\
c &=& 2-a, \nonumber\\
d &=& 1-b,\label{eq:theta_class1}
\end{eqnarray}
with $\theta \in[0,\pi]$. With this parametrization of $\theta$,  we express the corresponding entanglement witness for this class I as $W_I(\theta)$. 
Here $\theta=0$ corresponds to $W[1,1,1,0]$  which is the generalization of the Choi EW from $M_3(\cc)$. Similarly,  for $\theta=\pi$, we get another Choi-like witness  $W[1,0,1,1]$. It is important to mention here that the witness $W[a,b,c,d]$ is decomposable only if $b=d$ \cite{JPA}, which is equivalent to $\theta = \pi/2$. To check the optimality of the entanglement witness $W_I(\theta)$ for $\theta \neq \frac \pi 2$, let us look for a family of vectors $|\psi_k \otimes \phi_k\rangle$ satisfying Eq.~(\ref{!}).
For that purpose, we introduce a vector $|\psi\>\in\cc^4$ given by
\begin{equation}
 |\psi\> = \sum_{k=0}^3 e^{i \alpha_k} |k\>,
\end{equation}
for arbitrary (real) phases $\alpha_k$, and we observe that
\begin{equation}
 \< \psi \otimes \psi^*|W[a,b,c,d]|\psi \otimes \psi^* \> = 0.
\end{equation}
One has the following 
\begin{Proposition}   \label{PROP}
\label{prop1}
The vectors $|\psi_k\rangle \ot |\psi^*_k\rangle \in \mathbb{C}^n \ot \mathbb{C}^n$ with
\begin{equation}
\label{form1}
    |\psi_k\> = \sum_{\ell=0}^{n-1} e^{i\nu_{k\ell}} |\ell\> \ ,
\end{equation}
  with  real $\nu_{kl}$, span $n^2-(n-1)$ dimensional space in $\mathbb{C}^n \ot \mathbb{C}^n$.
\end{Proposition}
For $n=4$, it gives therefore 13 vectors. To have a spanning property, one still needs three additional linearly independent vectors satisfying Eq.~(\ref{!}). Interestingly, for $\theta =0$ and $\theta=\pi$, i.e. for Choi-like witnesses, we found only these 13 vectors, that satisfy (\ref{!}).
This is in a full analogy with the property of Choi witnesses for $n=3$: in that case one has only $9^2-2=7$ linearly independent vectors \cite{Kye-PRA,Gniewko}. 
Now we study the region $\theta \in (0,\pi)$, for which $a < 1$ and $b>0$. Introducing the following vectors
\begin{eqnarray}
\psii{14}&=&\sqrt{\sin(\theta/2)}\,|0\rangle+ \sqrt{\cos(\theta/2)}\,\1,  \nonumber \\
\fif{14} &=& \sqrt{\cos(\theta/2)}\,|0\rangle+ \sqrt{\sin(\theta/2)}\,\1, \nonumber\\
\psii{15} &=& \sqrt{\sin(\theta/2)}\,\1+ \sqrt{\cos(\theta/2)}\,\2, \nonumber \\
\fif{15} &=& \sqrt{\cos(\theta/2)}\,\1+ \sqrt{\sin(\theta/2)}\,\2, \\
\psii{16} &=& \sqrt{\sin(\theta/2)}\,\2+ \sqrt{\cos(\theta/2)}\,\3, \nonumber \\
\fif{16} &=& \sqrt{\cos(\theta/2)}\,\2+ \sqrt{\sin(\theta/2)}\,\3, \nonumber
\end{eqnarray}
one can show that
$$   \< \psi_k \ot \varphi_k |W[a,b,c,d]|\psi_k \ot \varphi_k  \> = 0 , $$
for $k=14,15,16$. 
However,  these 16 vectors  span only 15 dimensional space in $\mathbb{C}^4 \ot \mathbb{C}^4$.
There exists no other linearly independent vectors in this subspace. We will show it by subtracting an amount of the  projector onto its orthogonal complement and still obtaining an entanglement witness. 
 Summarising: for  class I we have 15 linearly independent vectors for $0 < \theta < \pi$ and 13  linearly independent vectors for $\theta = 0, \pi$. Recall that, if $n=3$, then for the Choi witness, we have only 7 ($=n^2-n+1$) linearly independent vectors.

\subsection{Optimization for the class I}

It turns out that EWs from the class I are not optimal.  Note, that $W_I(\pi/2)$ is decomposable. Precisely, for this case, we obtain the following decomposition
\begin{equation}
\label{DEC}
  W_I\big(\frac \pi 2\big) =  W\big[\frac 12,\frac 12,\frac 32,\frac 12\big] =  2P+ A^\Gamma ,
\end{equation}
where
$P=|\Psi\>\<\Psi|$  is a rank-1 projector onto the maximally entangled state in $\cc^4 \otimes \cc^4$ with
\begin{equation}
\label{psi}
    |\Psi\> = \frac 12 \sum_{j=0}^3 (-1)^{j+1} |j \otimes j\>,
\end{equation}
and $A$ is a positive definite matrix
\begin{widetext}
\begin{equation}
A= \frac 12
\left(
\begin{array}{cccc|cccc|cccc|cccc}
 . & . & . & . & . & . & . & . & . & . & . & . & . & . & . & . \\
 . & 1 & . & . & -1 & . & . & . & . & . & . & . & . & . & . & . \\
 . & . & 3 & . & . & . & . & . & -3 & . & . & . & . & . & . & . \\
 . & . & . & 1 & . & . & . & . & . & . & . & . & -1 & . & . & . \\
 \hline . & -1 & . & . & 1 & . & . & . & . & . & . & . & . & . & . & . \\
 . & . & . & . & . & . & . & . & . & . & . & . & . & . & . & . \\
 . & . & . & . & . & . & 1 & . & . & -1 & . & . & . & . & . & . \\
 . & . & . & . & . & . & . & 3 & . & . & . & . & . & -3 & . & . \\
 \hline . & . & -3 & . & . & . & . & . & 3 & . & . & . & . & . & . & . \\
 . & . & . & . & . & . & -1 & . & . & 1 & . & . & . & . & . & . \\
 . & . & . & . & . & . & . & . & . & . & . & . & . & . & . & . \\
 . & . & . & . & . & . & . & . & . & . & . & 1 & . & . & -1 & . \\
 \hline . & . & . & -1 & . & . & . & . & . & . & . & . & 1 & . & . & . \\
 . & . & . & . & . & . & . & -3 & . & . & . & . & . & 3 & . & . \\
 . & . & . & . & . & . & . & . & . & . & . & -1 & . & . & 1 & . \\
 . & . & . & . & . & . & . & . & . & . & . & . & . & . & . & . \\
\end{array}
\right) ,
\end{equation}
\end{widetext}
where we replaced all zeros by dots. It is therefore clear that $W_I(\frac \pi 2)$ is not optimal since we can subtract a positive operator $2P$, that is,
$$W_I\big(\frac \pi 2\big) - 2P = A^\Gamma, $$
is an EW. Let us  observe that $|\Psi\>$ defined in (\ref{psi}) is orthogonal to 15-dimensional subspace spanned by
$$ \{ |\psi_k\> \otimes |\psi_k^*\> , \  |\psi_\ell\> \otimes |\varphi_\ell\> \}, $$
for $k=1,\ldots,13$ and $\ell=14,15,16$. Hence, following Refs. \cite{Lew,Lew2} one may try to optimise $W(\theta)$ subtracting a fraction of the projector $P$.
\begin{thm}
\label{TH1}
The following operator
\begin{equation}
    {W_I}'(\theta) = W_I(\theta) - \lambda P,
\end{equation}
with $P=|\Psi\>\<\Psi|$ satisfying Eq.~(\ref{psi}) and $\theta \in [0,\pi]$,
is an entanglement witness if and only if $\lambda\leq 2$. Moreover, for $\lambda=2$ the witness is optimal. 
\end{thm}
The proof is provided in Appendix~\ref{optimal_proof}.  
Interestingly, the optimal witness $ {W}_{\rm opt}(\theta) = W_I(\theta) - 2 P$ has the spanning property for $\theta\in (0,\pi)$. Indeed, we provide the full set of spanning vectors given in Eqs.~(\ref{ani3})-(\ref{ani7}) in Appendix~\ref{optimal_proof}.

\subsection{Class II}
In class II,  obeying the condition~(\ref{II}), we have the following parametrization
\begin{eqnarray}
a &= & \frac{1}{2} (1+\cos\theta), \nonumber\\
b &=& \frac{1}{2} (2-\sin\theta), \nonumber\\
c &=& 1-a,\nonumber\\
d &=& 2-b,\label{eq:theta_class2}
\end{eqnarray}
for $\theta\in [0,\pi]$. 
Similarly like class I,  we express the corresponding entanglement witness with $\theta$ parametrization for this class II as $W_{II}(\theta)$.
Note that for $\theta=\pi$, we find
\begin{equation}
    W_{II}(\pi) = W[0,1,1,1] ,
\end{equation}
which recovers the EW corresponding to the reduction map. This witness is optimal having the spanning property, that is, 
$\langle \phi \otimes \phi^*|W(\pi)|\phi \otimes \phi^*\rangle=0$ for any $|\phi\rangle \in \mathbb{C}^4$ and these vectors span the whole 16 dimensional space.
 For $\theta=0$, we get another decomposable witness
\begin{equation}
    W_{II}(0) = W[1,1,0,1] ,
\end{equation}
with  the following decomposition
\begin{equation}
     W[1,1,0,1] = B^\Gamma + 2(P_1 + P_2) ,
\end{equation}
where
\begin{widetext}
\begin{equation}
B=
\left(
\begin{array}{cccc|cccc|cccc|cccc}
 . & . & . & . & . & . & . & . & . & . & . & . & . & . & . & . \\
 . & 1 & . & . & -1 & . & . & . & . & . & . & . & . & . & . & . \\
 . & . & . & . & . & . & . & . & . & . & . & . & . & . & . & . \\
 . & . & . & 1 & . & . & . & . & . & . & . & . & -1 & . & . & . \\
 \hline . & -1 & . & . & 1 & . & . & . & . & . & . & . & . & . & . & . \\
 . & . & . & . & . & . & . & . & . & . & . & . & . & . & . & . \\
 . & . & . & . & . & . & 1 & . & . & -1 & . & . & . & . & . & . \\
 . & . & . & . & . & . & . & . & . & . & . & . & . & . & . & . \\
 \hline . & . & . & . & . & . & . & . & . & . & . & . & . & . & . & . \\
 . & . & . & . & . & . & -1 & . & . & 1 & . & . & . & . & . & . \\
 . & . & . & . & . & . & . & . & . & . & . & . & . & . & . & . \\
 . & . & . & . & . & . & . & . & . & . & . & 1 & . & . & -1 & . \\
 \hline . & . & . & -1 & . & . & . & . & . & . & . & . & 1 & . & . & . \\
 . & . & . & . & . & . & . & . & . & . & . & . & . & . & . & . \\
 . & . & . & . & . & . & . & . & . & . & . & -1 & . & . & 1 & . \\
 . & . & . & . & . & . & . & . & . & . & . & . & . & . & . & . \\
\end{array}
\right),
\end{equation}
\end{widetext}
and $P_k = |\Psi_k\>\<\Psi_k|$, $k=1, 2$ with
\begin{eqnarray}
|\Psi_1\> &=& \frac{1}{\sqrt{2}} ( |0 \otimes 0\> - |2 \otimes 2\>, \label{p1_form} \\
|\Psi_2\> &=& \frac{1}{\sqrt{2}} ( |1 \otimes 1\> - |3 \otimes 3\>. \label{p2_form}
\end{eqnarray}
Hence $W[1,1,0,1]$ is not optimal. 

Next we investigate what happens when $\theta > 0$.
\begin{Proposition}  \label{PII}
The expectation value of $W_{II}(\theta)$ vanishes on a product vector iff it is of the form 
$|\psi^* \otimes \psi\rangle$, where
\begin{align}
\label{newcon}
    |\psi_0| & = |\psi_2|, & |\psi_1| & = |\psi_3|.
    \end{align}
\end{Proposition}
We provide prove this proposition in Appendix~\ref{appenC}. Moreover, we show  that
such vectors span 14 dimensional subspace of $\mathbb{C}^4 \otimes \mathbb{C}^4$.

\begin{thm} 
\label{TH2}
The witness $W_{II}(\theta)$ with $\theta \in (0,\pi)$ is optimal (with only 14 vectors satisfying (\ref{!})).
\end{thm}
We provide a proof of this theorem in Appendix~\ref{optimal_proof2}. 
Summarising: the class II consists of optimal EWs $W_{II}(\theta)$ for $\theta \in (0,\pi]$, where we have the spanning property only for $\theta=\pi$.

\section{Conclusions}
\label{conclude}
In this paper, we  considered two 1-parameter classes of  the entanglement witnesses in $\mathbb{C}^4 \otimes \mathbb{C}^4$, that are diagonal in the Bell basis. Additionally, observables from these classes are covariant w.r.t. maximal commutative subgroup of $U(4)$. The investigated classes are natural extension of a well-studied construction in $\mathbb{C}^3 \otimes \mathbb{C}^3$ of optimal witnesses described by a single parameter. Interestingly, the latter case contains paradigmatic examples of EWs, that are indecomposable Choi witness and the witness corresponding to the reduction map. Now in $\mathbb{C}^4 \otimes \mathbb{C}^4$, the situation is different: instead of a single class, one has two classes which display distinct properties, and can serve as a playground for investigation of various features of entanglement witnesses. 

Class I contains only EWs  which are not optimal. This shows that a straightforward generalization of the Choi witness from $n=3$ to $n=4$ does not preserve optimality. Let us recall that for $n=3$ the Choi  witness is not only optimal but even extremal \cite{choi1,Kye-PRA}. Moreover, all EWs from this class posses 15 
linearly independent vectors satisfying condition (\ref{!}). The only exception is provided by Choi-like witnesses for which one has only 13 vectors. Following the optimization technique developed in Refs. \cite{Lew,Lew2}, we have shown that all EWs from class I can be optimized by subtracting a single projector. Interestingly, the optimized entanglement witnesses possess the spanning property (again with an exception of Choi-like witnesses for which we have now 14 vectors satisfying (\ref{!})).

Class II contains optimal EWs (with one exception $W_{II}(0)$ i.e. at $\theta=0$). However, these EWs do not have the spanning property (again with one exception $W(\pi)$ corresponding to the reduction map). This result is quite unexpected since it shows that in the limit $\theta \to 0$, one obtains non-optimal witness $W_{II}(0)$ out of optimal EWs $W_{II}(\theta)$.

It should be stressed that checking for optimality of a given EW is in general a difficult problem. 
This analysis  sheds a new light into the structure of optimal entanglement witnesses. We show that two classes of EWs displaying the same symmetry posses very different properties: {one is optimal and the other is not}. Moreover, due to the symmetry of the problem we were able to performed the full optimization procedure \cite{Lew,Lew2}. It would be interesting to find general characterization for arbitrary $n$. Another interesting problem is the issue of extremality. For $n=3$ it is known that the considered class is already extremal. For $n>3$ the problem is open.


\section*{Acknowledgements}
The work was supported by the Polish National Science Centre project No. 2018/30/A/ST2/00837. 
FW is thankful for support from NASA Academic Mission Services, Contract No. NNA16BD14C.
\\
\appendix

\section{Proof  of Theorem \ref{TH1}}
\label{optimal_proof}
Due to the Choi–Jamiołkowski isomorphism, any entanglement witness $W$ in 
$\mathbb{C}^n \otimes \mathbb{C}^n$ corresponds to a positive map 
$\Phi:M_n(\mathbb{C}) \otimes M_n(\mathbb{C})$  via the following relation
\begin{equation}
W=\sum_{i,j=0}^{n-1} e_{ij} \otimes \Phi(e_{ij}),
\end{equation}
where $e_{ij}=|e_i \rangle \langle e_j|$,  $\{e_0, e_1,\ldots,e_{n-1}\}$ denotes an orthonormal basis in $\mathbb{C}^n$.
The map corresponding to Eq.~(\ref{W!}) has the following form
\begin{eqnarray}
\Phi(e_{ii})=\sum_{j=0}^{n-1} a_{ij} e_{jj}, \nonumber\\
\Phi(e_{ij})=e_{ij}, ~i \neq j
\end{eqnarray}
where $a_{ij}=\alpha_{i-j} \geq 0$. One can easily find the inverse relation, which is
\begin{equation}
\label{dekhi1}
\Phi(|\chi\rangle \langle\chi|)=\Tr_1((|\chi\rangle \langle \chi|)^T \otimes \mathbb{I}_n~W),
\end{equation}
where the transposition is performed with respect to $\{e_0, e_1,\ldots,e_{n-1}\}$.
Now, to show the optimality of the Theorem \ref{TH1}, i.e. the entanglement witness $W_I'(\theta)=W_I(\theta)-2 P$ in first class, our idea is to find the 16 linearly independent vectors satisfying the spanning criteria for the EW $W_I'(\theta)$.  For that purpose,  we  act with the corresponding map $\Phi$ on an arbitrary vector $|\psi\rangle =\{\psi_0,\psi_1,\psi_2,\psi_3\} \in \mathbb{C}^4$ and  get 
\begin{widetext}
\begin{equation}
 \Phi(|\psi\rangle\langle\psi|) = \mathrm{diag} \{ y_0, y_1, y_2, y_3 \} - | \psi \rangle \langle \psi | -
 2 D \circ | \psi \rangle \langle \psi |
    \stackrel{df}{=} A - B - C,
\end{equation}
  where $\circ$ denotes the Hadamard product,
\begin{equation}
    D= \left(\begin{array}{cccc} 1 & -1 & 1 & -1 \\ -1 & 1 & -1 & 1 \\ 1 & -1 & 1 & -1 \\ -1 & 1 & -1 & 1 \end{array} \right) ,
\end{equation}  
  and 
  \begin{align}
    y_0 = \left(1 + \frac{2-\sin\theta}2\right) |\psi_0|^2 + \frac{1+\cos\theta}2 |\psi_1|^2 + \frac{2+\sin\theta}2 |\psi_2|^2 + \frac{1-\cos\theta}2 |\psi_3|^2, \\
    y_1 = \left(1 + \frac{2-\sin\theta}2\right) |\psi_1|^2 + \frac{1+\cos\theta}2 |\psi_2|^2 + \frac{2+\sin\theta}2 |\psi_3|^2 + \frac{1-\cos\theta}2 |\psi_0|^2, \\
    y_2 = \left(1 + \frac{2-\sin\theta}2\right) |\psi_2|^2 + \frac{1+\cos\theta}2 |\psi_3|^2 + \frac{2+\sin\theta}2 |\psi_0|^2 + \frac{1-\cos\theta}2 |\psi_1|^2, \\
    y_3 = \left(1 + \frac{2-\sin\theta}2\right) |\psi_3|^2 + \frac{1+\cos\theta}2 |\psi_0|^2 + \frac{2+\sin\theta}2 |\psi_1|^2 + \frac{1-\cos\theta}2 |\psi_2|^2.
  \end{align}
    Its determinant is equal to
  \begin{align}
    \det[\Phi(|\psi\rangle\langle\psi|)]
    & = \det[A_0 | A_1 | A_2 | A_3] - \det[B_0 | A_1 | A_2 | A_3] - \dots - \det[A_0 | A_1 | A_2 | B_3] 			
     - \det[C_0 | A_1 | A_2 | A_3] \nonumber\\
     & - \dots - \det[A_0 | A_1 | A_2 | C_3] + \det[B_0 | C_1 | A_2 | A_3] + \det[C_0 | B_1 | A_2 | A_3] + \dots \nonumber\\
     & + \det[C_0 | A_1 | A_2 | B_3] + \det[B_0 | A_1 | A_2 | C_3].
  \end{align}
  \end{widetext}
This determinant is a multilinear function on columns, hence we obtain $4^3$ summands, but most of them are zero -  in each summand, there can be at most one column from $B$ and at most one column from $C$ (each two columns of $B$ are linearly dependent and then the summand would be zero, the same holds for $C$). If one column from $B$ and one column from $C$ enter the summand, they have to be neighbours to produce a non-zero summand.
  Proceeding in this way, we get
   \begin{widetext}
  \begin{align}
    \det[\Phi(|\psi\rangle\langle\psi|)]
     &= y_0y_1y_2y_3 - \frac 32 |\psi_0|^2 y_1y_2y_3 - \dots - \frac 32 |\psi_1|^2 y_2y_3y_0 \nonumber\\
     &+ 2 |\psi_0|^2 |\psi_1|^2 y_2 y_3 + 2 |\psi_1|^2 |\psi_2|^2 y_3 y_0 + 2 |\psi_2|^2 |\psi_3|^2 y_0 y_1 + 2 |\psi_3|^2 |\psi_0|^2 y_1 y_2 \nonumber\\
    &= y_0y_1y_2y_3 \left[ 1 - \frac 32 \sum_{i=0}^3 \frac{|\psi_i|^2}{y_i} + 2 \left( \frac{|\psi_0|^2}{y_0} + \frac{|\psi_2|^2}{y_2}\right) \left( \frac{|\psi_1|^2}{y_1} + \frac{|\psi_3|^2}{y_3}\right) \right].
  \end{align}
   \end{widetext}
 Now, we will show that the second factor in the above is non-negative. Denoting
  \begin{align}
   z_0 & = \frac{|\psi_0|^2}{y_0} + \frac{|\psi_2|^2}{y_2}, \\
   z_1 & = \frac{|\psi_1|^2}{y_1} + \frac{|\psi_3|^2}{y_3}.
  \end{align}
  We have the following condition
  \begin{eqnarray}
  \label{pos1}
    1 - \frac 32 (z_0 + z_1) + 2z_0z_1 \ge 0 \nonumber\\
     \Rightarrow ( 4z_0 - 3 ) ( 4z_1 - 3 ) \ge 1,
  \end{eqnarray}
  where
  \begin{widetext}
    \begin{align}
    4 z_0 - 3 = - \frac{(x_{0+} + 3x_{1+})(3x_{0+} + x_{1+})+((5+3s)x_{0-} - 3c x_{1-})((1-s)x_{0-} + cx_{1-})}{\left( 3 x_{0+} + x_{1+} \right)^2 - \left( (1-s) x_{0-} + c x_{1-} \right)^2},  \label{ani1}\\
    4 z_1 - 3 = - \frac{(x_{0+} + 3x_{1+})(3x_{0+} + x_{1+})+((5+3s)x_{1-} + 3c x_{0-})((1-s)x_{1-} - cx_{0-})}{\left( 3 x_{1+} + x_{0+} \right)^2 - \left( (1-s) x_{1-} - c x_{0-} \right)^2}. \label{ani2}
  \end{align}
  \end{widetext}
  Note that we denote here  $\sin\theta$ and $\cos\theta$ as $s$ and   $c$ respectively, and also
  \begin{align}
    x_{0\pm} = |\psi_0|^2 \pm |\psi_2|^2, \label{x0p} \\
    x_{1\pm} = |\psi_1|^2 \pm |\psi_3|^2. \label{x1p}
  \end{align}
  Using Eqs.~(\ref{ani1}) and (\ref{ani2}) in (\ref{pos1}), we obtain
  \begin{widetext}
  \begin{align}
    &2(1-s)(x_{0+} + 3x_{1+})(3x_{0+} + x_{1+})(x_{0-}^2 + x_{1-}^2) +& \nonumber\\
    &(1-s)\big(c(x_{1-}^2 - x_{0-}^2) - 2s x_{0-} x_{1-}\big)
    \big((5+3s)x_{0-} - 3c x_{1-}\big)
    \big((5+3s)x_{1-} + 3c x_{0-}\big) &\nonumber \\
    &\ge
    -
    (1-s)
    \left( 3 x_{0+} + x_{1+} \right)^2
    \left( (1-s) x_{1-}^2 - 2c x_{0-}x_{1-} + (1+s) x_{0-}^2 \right)^2 &\nonumber \\
    &-
    (1-s)
    \left( 3 x_{1+} + x_{0+} \right)^2
    \left( (1-s) x_{0-}^2 + 2c x_{0-}x_{1-} + (1+s) x_{1-}^2 \right)^2 &\nonumber \\
    &+
    (1-s)^2 \left( c(x_{1-}^2 -x_{0-}^2) - 2s x_{1-} x_{0-} \right)^2,&
      \end{align}
which implies
\begin{align}
    &2 (x_{0-}^2 + x_{1-}^2) (x_{0+} + x_{1+})^2
    + s (x_{0-}^2 - x_{1-}^2) (x_{0+}^2 - x_{1+}^2)
    - c ( 2 x_{0-} x_{1-} ) (x_{0+}^2-x_{1+}^2)&
    \nonumber \\
    &+ c (1+4s+2s^2) ( 2 x_{0-} x_{1-}) (x_{1-}^2-x_{0-}^2)
    - s (1+s)^2 (2x_{0-} x_{1-})^2
    - c^2 (2+s) (x_{1-}^2 - x_{0-}^2)^2 \ge 0,& \nonumber\\
     &\Rightarrow   2 (x_{0-}^2 + x_{1-}^2) (x_{0+} + x_{1+})^2
    - (s (x_{1-}^2 - x_{0-}^2) + c \cdot 2x_{0-}x_{1-}) (x_{0+}^2-x_{1+}^2)&
    \nonumber \\
    &- (c (x_{1-}^2 - x_{0-}^2) - s \cdot 2x_{0-}x_{1-}) (c(2+s)(x_{1-}^2 - x_{0-}^2) - (1+s)^2 \cdot 2x_{0-}x_{1-})
    \ge 0,&
    \label{last_nonpolar}
    \end{align}
  \end{widetext}
  Clearly here $x_{0+}, x_{1+} \ge 0$ and  $x_{0-} \in [ -x_{0+}, x_{0+} ]$, $ x_{1-} \in [ -x_{1+}, x_{1+} ]$. Now we convert our system into the polar coordinates and will rewrite the above equation in these new coordinates.

   \begin{figure}[t]
  \begin{tikzpicture}[>=latex]
    \draw (-3,-2) -- (-3,2) -- (3,2) -- (3,-2) -- cycle;
    \draw[->] (0,-3) -- (0,3);
    \draw[->] (-4,0) -- (4,0);
    \fill (-3,0) circle (.05);
    \fill (3,0) circle (.05);
    \fill (0,-2) circle (.05);
    \fill (0,2) circle (.05);
    \node at (4,-.3) {$x_{0-}$};
    \node at (2.7,-.3) {$x_{0+}$};
    \node at (-2.5,-.3) {$-x_{0+}$};
    \node at (.4,2.8) {$x_{1-}$};
    \node at (-.4,1.8) {$x_{1+}$};
    \node at (-.5,-1.8) {$-x_{1+}$};
    \draw (0,0) -- (4,4);
    \draw[dashed] (3,2) -- (3,4);
    \fill (2,2) circle (.05);
    \fill (3,3) circle (.05);
    \fill (0,0) circle (.05);
    \draw[->] (1,0) arc (0:45:1);
    \node at (.5,1) {$R$};
    \node at (.6,.3) {$\phi$};
  \end{tikzpicture}
 \caption{The polar coordinates representation. Here, $\phi \in [0,2\pi)$. }
  \end{figure}
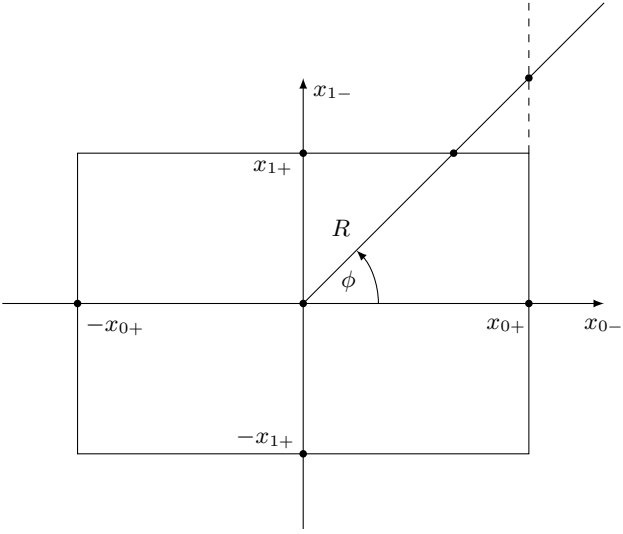

 In the new coordinate system,  $R \in [0, \min \big\{ \frac{x_{0+}}{|\cos\phi|} , \frac{x_{1+}}{|\sin\phi|} \big\} ]$, and therefore the inequality (\ref{last_nonpolar}) can be written as
  \begin{eqnarray}
    R^2 \left(
    2 (x_{0+} + x_{1+})^2
    - \sin (2\phi - \theta) (x_{0+}^2-x_{1+}^2)
    \right) \nonumber\\
    - R^4 \cos(2\phi-\theta) (\sin 2\phi + (2+\sin\theta)\cos(2\phi-\theta)) 
    \ge 0,\nonumber\\
  \end{eqnarray}
  and this equation has to be satisfied for all $R$ mentioned in the above range. Hence
  \begin{widetext}
  \begin{align}
    \cos(2\phi-\theta) \big(\sin 2\phi + (2+\sin\theta)\cos(2\phi-\theta)\big)
    \min \left\{ \frac{x_{1+}^2}{\sin^2\phi} , \frac{x_{0+}^2}{\cos^2\phi} \right\}
    \le 2 (x_{0+} + x_{1+})^2 - \sin (2\phi - \theta) (x_{0+}^2-x_{1+}^2).
  \end{align}
  \end{widetext}
  First, we consider the scenario when $\frac{x_{1+}^2}{\sin^2\phi} \le \frac{x_{0+}^2}{\cos^2\phi}$. Therefore,
  \begin{align}
    &\cos(2\phi-\theta) \big(\sin 2\phi + (2+\sin\theta)\cos(2\phi-\theta)\big) x_{1+}^2& \nonumber\\
    &\le \sin^2\phi \left( 2 (x_{0+} + x_{1+})^2 - \sin (2\phi - \theta) (x_{0+}^2-x_{1+}^2) \right).& \nonumber\\
  \end{align}
  Ordering the terms, we get
  \begin{align}
    &x_{0+}^2  \sin^2\phi \left( 2  - \sin (2\phi - \theta) \right)
    + 4 \sin^2\phi \ x_{0+}x_{1+} +& \nonumber\\
    & x_{1+}^2  \big[
    \sin^2\phi
    (2 + \sin (2\phi - \theta))
    - \cos(2\phi-\theta) \sin 2\phi& \nonumber\\
    &- (2+\sin\theta)\cos^2(2\phi-\theta)\big]  \ge 0.&
  \end{align}

  It is equivalent to say that the quadratic function
  \begin{align}
    \frac{x_{0+}}{x_{1+}} \mapsto
    &
    \sin^2\phi \left( 2  - \sin (2\phi - \theta) \right) \left( \frac{x_{0+}}{x_{1+}} \right)^2+
    \nonumber\\ &
     4 \sin^2\phi \frac{x_{0+}}{x_{1+}}
    +
    \big(
    \sin^2\phi
    (2 + \sin (2\phi - \theta) )  -\nonumber\\
    &
    \cos(2\phi-\theta) \sin 2\phi
     - (2+\sin\theta)\cos^2(2\phi-\theta)
    \big),
  \end{align}
  has to be positive in the range  $[|\cot \phi|,\infty)$. As the function is increasing for positive arguments, it is equivalent to check its positivity at the end point, i.e at the point $|\cot \phi|$
  \begin{align}
  &  \left( 2  - \sin (2\phi - \theta) \right) \cos^2\phi
    + 4 |\sin\phi\cos\phi|&
    \nonumber \\
   & +
    \sin^2\phi
    (2 + \sin (2\phi - \theta))
    - \cos(2\phi-\theta) \sin 2\phi &\nonumber\\
   & - (2+\sin\theta)\cos^2(2\phi-\theta)
    \ge
    0.&
  \end{align}
  After simplification, we arrive at
  \begin{eqnarray}
    2 + 2|\sin2\phi|-\sin(4\phi-\theta)-(2+\sin\theta)\cos^2(2\phi-\theta)  \nonumber\\
     \ge 0.  \nonumber\\
    \label{main_ineq}
  \end{eqnarray}
  To show that this inequality holds, it suffices to consider $\phi \in [0,\pi/2]$ due to periodicity of the LHS. 
And therefore one can remove the modulus  from the above inequality.   
  Decomposing the coeffiecients of the above inequality  in the following way
  \begin{align}
  & \sin(4\phi-\theta)  = \sin(2\phi) \cos(2\phi-\theta) +&\nonumber\\
    &\cos(2\phi) \sin(2\phi-\theta),& \\
   &\sin(\theta)  = \sin(2\phi) \cos(2\phi-\theta) -& \nonumber\\
 &  \cos(2\phi) \sin(2\phi-\theta),&
  \end{align}
  and organizing terms, one can rewrite the above inequality~(\ref{main_ineq}) as:
  \begin{eqnarray}
    \sin^2(2\phi-\theta) \left( 2-\cos(2\phi)\sin(2\phi-\theta) \right) + \nonumber\\
    \sin 2\phi \left( 2 - \cos (2\phi-\theta) - \cos^3 (2\phi-\theta) \right) \ge 0. \nonumber\\
        \label{main_ineq1}
  \end{eqnarray}

  This inequality obviously holds, as all factors in both summands are non-negative. One can clearly see that the inequality is saturated iff $2\phi = \theta$, hence it is saturated for four values of $\phi$ if $\theta \in [0 ,\pi]$ due to periodicity. Additionally, this inequality will never be saturated if $\theta \in (\pi,2\pi)$ because, in this range of $\theta$, $\phi>\pi/2$ which implies we are already out of our domain. 

Now, we consider the other scenario i.e. $\frac{x_{1+}^2}{\sin^2\phi} \ge \frac{x_{0+}^2}{\cos^2\phi}$. Then
  \begin{align}
    \cos(2\phi-\theta) \big(\sin 2\phi + (2+\sin\theta)\cos(2\phi-\theta)\big) x_{0+}^2
    \le \nonumber\\
     \cos^2\phi \left( 2 (x_{0+} + x_{1+})^2 - \sin (2\phi - \theta) (x_{0+}^2-x_{1+}^2) \right). \nonumber\\
  \end{align}
  By ordering the terms, we obtain
  \begin{eqnarray}
 &&   \cos^2\phi \big( 2  + \sin (2\phi - \theta) \big) x_{1+}^2
    + 4 \cos^2\phi \ x_{0+}x_{1+}
    \nonumber \\
    & + &
    \big( \cos^2\phi
    (2 - \sin (2\phi - \theta) )
    - \cos(2\phi-\theta) \sin 2\phi \nonumber\\
& - &  (2+\sin\theta)\cos^2(2\phi-\theta) \big)
    x_{0+}^2
    \ge   0.
  \end{eqnarray}
  It is equivalent to say that the quadratic function:
  \begin{eqnarray}
    \frac{x_{1+}}{x_{0+}} & \mapsto &
    \cos^2\phi \left( 2  + \sin (2\phi - \theta) \right) \left( \frac{x_{1+}}{x_{0+}} \right)^2 \nonumber\\
    &  + & 4 \cos^2\phi \frac{x_{1+}}{x_{0+}}
    + 
    \big(
    \cos^2\phi
    (2 - \sin (2\phi - \theta) ) \nonumber\\
     &  - &  \cos(2\phi-\theta) \sin 2\phi 
    -  (2+\sin\theta)\cos^2(2\phi-\theta)
    \big), \nonumber\\
  \end{eqnarray}
  has to be positive in the range $[|\tan \phi|,\infty)$. As the function is increasing for positive arguments, it is equivalent to check its positivity at the point $|\tan \phi|$:
  \begin{align}
   & \sin^2\phi \left( 2  + \sin (2\phi - \theta) \right)
    + 4 |\cos\phi| |\sin\phi|& \nonumber \\
   & +  \cos^2\phi
    (2 - \sin (2\phi - \theta) )
    - \cos(2\phi-\theta) \sin 2\phi & \nonumber\\
  &  - (2+\sin\theta)\cos^2(2\phi-\theta)
    \ge   0. &\nonumber\\
  \end{align}
  Simplifying the above inequality, we obtain again the same inequality (\ref{main_ineq}), which has been proven already.

  The inequality (\ref{main_ineq1}) is saturated if $\theta \in [0,\pi]$ and $\phi = \frac\theta 2 + k \frac\pi 2$. We have obtained (\ref{main_ineq1}) putting $x_{0+} = |\sin\phi|$, $x_{1+} = |\cos\phi|$, and hence $R=1$. The inequality (\ref{last_nonpolar}) is satisfied in the following four points:
\begin{widetext}
  \begin{align}
    x_{0+} & = \cos\frac\theta 2 ,& x_{1+} & = \sin\frac\theta 2 ,& x_{0-} & = \cos \frac \theta 2 ,& x_{1-} & = \sin \frac \theta 2 ,\\
    x_{0+} & = \sin\frac\theta 2 & x_{1+}, & = \cos\frac\theta 2, & x_{0-} & = -\sin \frac \theta 2, & x_{1-} & = \cos \frac \theta 2, \\
    x_{0+} & = \cos\frac\theta 2, & x_{1+} & = \sin\frac\theta 2, & x_{0-} & = -\cos \frac \theta 2, & x_{1-} & = -\sin \frac \theta 2, \\
    x_{0+} & = \sin\frac\theta 2, & x_{1+} & = \cos\frac\theta 2, & x_{0-} & = \sin \frac \theta 2 & x_{1-} & = -\cos \frac \theta 2.
  \end{align}
  Additional solutions are for $R=0$. Then $x_{0-}=x_{1-}=0$ and $x_{0+},x_{1+}$ are arbitrary. The additional solution exists for the whole range of $\theta$. It corresponds to
  \begin{align}
    |\psi_0|^2 & = \cos\frac\theta 2, & |\psi_1|^2 & = \sin\frac\theta 2, & |\psi_2|^2 & = 0, & |\psi_3|^2 & = 0,\label{ani3} \\
    |\psi_0|^2 & = 0, & |\psi_1|^2 & = \cos\frac\theta 2, & |\psi_2|^2 & = \sin \frac \theta 2, & |\psi_3|^2 & = 0,\label{ani4} \\
    |\psi_0|^2 & = 0, & |\psi_1|^2 & = 0, & |\psi_2|^2 & = \cos \frac \theta 2, & |\psi_3|^2 & = \sin \frac \theta 2,\label{ani5} \\
    |\psi_0|^2 & = \sin\frac\theta 2, & |\psi_1|^2 & = 0, & |\psi_2|^2 & = 0, & |\psi_3|^2 & = \cos \frac \theta 2,\label{ani6} \\
    |\psi_0|^2 & = t, & |\psi_1|^2 & = 1-t, & |\psi_2|^2 & = t, & |\psi_3|^2 & = 1-t.\label{ani7}
  \end{align}
  For such vectors $|\psi\rangle$ with the arbitrary choice of phases, the determinant of $\Phi(|\psi\rangle\langle\psi|)$ is zero. Therefore,  the kernel of $\Phi(|\psi\rangle\langle\psi|)$ is  spanned by respectively:
  \begin{align}
    &\ker \Phi\left(\left[
    \begin{array}{l} \sqrt{\cos\frac\theta 2} \\ \sqrt{\sin\frac\theta 2} e^{i\phi} \\ 0 \\ 0 \end{array}
    \right] \left[
    \begin{array}{llll} \sqrt{\cos\frac\theta 2} & \sqrt{\sin\frac\theta 2} e^{-i\phi} & 0 & 0 \end{array}
    \right] \right)
    = \mathrm{span}
    \left[
    \begin{array}{l} \sqrt{\sin\frac\theta 2} e^{i\phi} \\ \sqrt{\cos\frac\theta 2} \\ 0 \\ 0 \end{array}
    \right],
    \\
    & \ker \Phi\left(\left[
    \begin{array}{l} 0 \\ \sqrt{\cos\frac\theta 2} \\ \sqrt{\sin\frac\theta 2} e^{i\phi}  \\ 0 \end{array}
    \right] \left[
    \begin{array}{llll} 0 & \sqrt{\cos\frac\theta 2} & \sqrt{\sin\frac\theta 2} e^{-i\phi}  & 0 \end{array}
    \right] \right)
    = \mathrm{span}
    \left[
    \begin{array}{l} 0 \\ \sqrt{\sin\frac\theta 2} e^{i\phi} \\ \sqrt{\cos\frac\theta 2} \\ 0 \end{array}
    \right],
    \\
    & \ker \Phi\left(\left[
    \begin{array}{l} 0 \\ 0 \\ \sqrt{\cos\frac\theta 2} \\ \sqrt{\sin\frac\theta 2} e^{i\phi} \end{array}
    \right] \left[
    \begin{array}{llll} 0 & 0 & \sqrt{\cos\frac\theta 2} & \sqrt{\sin\frac\theta 2} e^{-i\phi} \end{array}
    \right] \right)
    = \mathrm{span}
    \left[
    \begin{array}{l} 0 \\ 0 \\ \sqrt{\sin\frac\theta 2} e^{i\phi} \\ \sqrt{\cos\frac\theta 2} \end{array}
    \right],   
     \\
     & \ker \Phi\left(\left[
    \begin{array}{l} \sqrt{\sin\frac\theta 2} e^{i\phi} \\ 0 \\ 0 \\ \sqrt{\cos\frac\theta 2} \end{array}
    \right] \left[
    \begin{array}{llll} \sqrt{\sin\frac\theta 2} e^{-i\phi} & 0 & 0 & \sqrt{\cos\frac\theta 2} \end{array}
    \right] \right)
    = \mathrm{span}
    \left[
    \begin{array}{l} \sqrt{\cos\frac\theta 2} \\ 0 \\ 0 \\ \sqrt{\sin\frac\theta 2} e^{i\phi} \end{array}
    \right],
    \\
    &\ker \Phi\left(\left[
    \begin{array}{l} \sqrt{t} \\ \sqrt{1-t} e^{i\alpha} \\ \sqrt{t} e^{i\beta} \\ \sqrt{1-t} e^{i\gamma} \end{array}
    \right] \left[
    \begin{array}{llll} \sqrt{t} & \sqrt{1-t} e^{-i\alpha} & \sqrt{t} e^{-i\beta} & \sqrt{1-t} e^{-i\gamma} \end{array}
    \right] \right)
    = \mathrm{span}
    \left[
    \begin{array}{l} \sqrt{t} \\ \sqrt{1-t} e^{i\alpha} \\ \sqrt{t} e^{i\beta} \\ \sqrt{1-t} e^{i\gamma} \end{array}
    \right].
  \end{align}
  \end{widetext}

  Therefore we can say that for $\theta < \pi$, the last family~(\ref{ani7}) along with arbitrary phases
spans the 14-dimensional subspace  and 
    the additional solutions (\ref{ani3},\ref{ani4},\ref{ani5},\ref{ani6}) spans the whole 16-dimensional space. Hence our entanglement witness $W_I'(\theta)$ is optimal. It completes the proof. 

\section{Proof of Proposition \ref{PII}}

\label{appenC}

In this section, we will  prove the  Eq.~(\ref{newcon}). For that purpose, 
let us consider the local contraction of the witness $W_{II}(\theta)$ of the second class satisfying Eq.~(\ref{eq:theta_class2}) with the projector $|\psi\rangle \langle \psi|$ in the second subsystem,  and we get
\begin{eqnarray}
W_{\psi}(\theta)&=&\Tr_2\Big(\mathbb{I} \otimes |\psi\rangle \langle \psi| W_{II}(\theta)\Big) \nonumber\\
&=& \mathrm{diag} \{ y_0, y_1, y_2, y_3 \} - | \psi^* \rangle \langle \psi^* |,
\end{eqnarray}
where
\begin{widetext}
 \begin{align}
    y_0 = \frac{3+c}{2} |\psi_0|^2 + \frac{2-s}{2} |\psi_1|^2 + \frac{1-c}{2} |\psi_2|^2 + \frac{2+s}{2} |\psi_3|^2,\\
    y_1 = \frac{3+c}{2}|\psi_1|^2 + \frac{2-s}{2}|\psi_2|^2 + \frac{1-c}{2}  |\psi_3|^2 + \frac{2+s}{2} |\psi_0|^2,\\
    y_2 = \frac{3+c}{2} |\psi_2|^2 + \frac{2-s}{2} |\psi_3|^2 + \frac{1-c}{2}  |\psi_0|^2 + \frac{2+s}{2} |\psi_1|^2,\\
    y_3 =  \frac{3+c}{2} |\psi_3|^2 +\frac{2-s}{2} |\psi_0|^2 + \frac{1-c}{2} |\psi_1|^2 + \frac{2+s}{2}|\psi_2|^2 ,
  \end{align}
with $s := \sin \theta$ and $c := \cos\theta$. 
The determinant of  $W_{\psi}(\theta)$ is given by
\begin{align}
\det[W_{\psi}(\theta)]&=y_0y_1y_2y_3-|\psi_0|y_1 y_2 y_3-
|\psi_1| y_2 y_3 y_0-|\psi_2| y_3 y_0 y_1-|\psi_3| y_0 y_1 y_2 \nonumber\\
&=\Big[(X_0+X_1+X_2+X_3)^2-\big(\frac{1+c}{2} (X_0-X_2)-\frac{s}{2} (X_1-X_3)\big)^2\Big]\cdot 
\Big[(X_0+X_1+X_2+X_3)^2-\nonumber\\
 &\big(\frac{1+c}{2} (X_1-X_3)+\frac{s}{2} (X_0-X_2)\big)^2 \Big]-
\Big[(X_0+X_1+X_2+X_3)^2-\big(\frac{1+c}{2} (X_0-X_2)-\nonumber\\
&\frac{s}{2} (X_1-X_3)\big)^2\Big] \cdot
 \Big[(X_1+X_3)^2-\frac{1+c}{2} (X_1-X_3)^2+(X_1+X_3) (X_0+X_2)-\nonumber\\
 &\frac{s}{2} (X_1-X_3) (X_0-X_2) \Big]-
\Big[(X_0+X_1+X_2+X_3)^2-\big(\frac{1+c}{2} (X_1-X_3)+
\frac{s}{2} (X_0-X_2)\big)^2\Big] \cdot \nonumber\\ &\Big[(X_0+X_2)^2-\frac{1+c}{2} (X_0-X_2)^2+(X_1+X_3) (X_0+X_2)+\frac{s}{2} (X_1-X_3) (X_0-X_2) \Big] \nonumber\\
 &= S_1 + S_2 + S_3 ,
 \end{align}
where 
\begin{align} 
 S_1 &= \big(\frac{1+c}{2}\big)^2 \Big(\frac{s}{2} (X_0-X_2)^2-\frac{s}{2} (X_1-X_3)^2+c (X_0-X_2)(X_1-X_3)  \Big)^2,\nonumber\\
 S_2 &= \Big(\frac{1+c}{2} (X_0-X_2)-\frac{s}{2} (X_1-X_3) \Big)^2 \cdot 
 \Big((X_1+X_3) (X_0+X_1+X_2+X_3)-\nonumber\\
 &(X_1-X_3) \big( \frac{1+c}{2} (X_1-X_3)+\frac{s}{2}(X_0-X_2)\big)  \Big), \nonumber\\
 S_3 &= \Big(\frac{1+c}{2} (X_1-X_3)-\frac{s}{2} (X_0-X_2) \Big)^2 \cdot \nonumber\\
&\Big((X_0+X_2) (X_0+X_1+X_2+X_3)-(X_0-X_2) \big( \frac{1+c}{2} (X_0-X_2)-\frac{s}{2}(X_1-X_3)\big)\Big). 
 \end{align}
\end{widetext}
We   introduce the following notation $X_i :=|\psi_i|^2$, for $i=0,1,2,3$. We observe that $S_1 \geq 0$. Moreover 
\begin{eqnarray}
\label{dp1}
|(X_1-X_3) \big( \frac{1+c}{2} (X_1-X_3)+\frac{s}{2}(X_0-X_2)\big)|  \nonumber\\    \leq (X_1+X_3) (X_0+X_1+X_2+X_3), \nonumber\\
\end{eqnarray}
hence $S_2 \geq 0$. Similarly, we show that $S_3 \geq 0$.

If $\theta=\pi$, then $\frac{1+c}{2}=\frac{s}{2}=0$, and the contraction is always singular, because then the witness corresponds to the reduction map. In other case, i.e. $\theta \in [0,\pi)$, the inequality (\ref{dp1}) is never saturated, and the only way to have zero  determinant is: $X_0=X_2$ and $X_1=X_3$. For such $|\psi\rangle$, we have $W_\psi(\theta)=\mathbb{I} |\psi|^2-|\psi^*\rangle \langle \psi^*|$ and its kernel is spanned by $\{|\psi^*\rangle\}$. Hence the vectors $|\psi^*\otimes \psi\rangle$ are the only product vectors for which the expectation value of $W_{II}(\theta)$  vanishes.

One can observe that the vectors in Eqs.~(\ref{p1_form}) and (\ref{p2_form}) are orthogonal to all $|\psi^*\otimes \psi\rangle$ satisfying (\ref{newcon}). Hence 
\begin{eqnarray}
\dim \{ \psi^* \otimes \psi:\langle \psi^* \otimes \psi|W_{II}(\theta)|\psi^* \otimes \psi \rangle=0 \nonumber\\
 \land~ |\psi_0|=|\psi_2| ~\land~ |\psi_1|=|\psi_3| \} \leq 14.
\end{eqnarray}
Taking random 14 vectors from the above subspace, one  easily checks that the above inequality is saturated.

\section{Proof  of Theorem \ref{TH2}}
\label{optimal_proof2}
  To show  optimality of the entanglement witness $W_{II}(\theta)$ in class II we show that 
  for any vector $|\Psi_{x,y}\> = x |\Psi_1\> + y |\Psi_2\>$, with $|\Psi_1\>$ and $|\Psi_2\>$ defined in (\ref{p1_form}) and (\ref{p2_form}), respectively, the following operator
  \begin{eqnarray}
  W_{II}(\theta) - \lambda |\Psi_{x,y}\>\<\Psi_{x,y}| ,
  \end{eqnarray}
  is not an EW, whenever $\lambda > 0$. Consider the corresponding linear map defined in Eq.~(\ref{dekhi1}), that is, 
  \begin{eqnarray}
  \Phi_{\theta,\lambda,x,y}(X) := \Phi_\theta(X) - \lambda D_{x,y} \circ X,
  \end{eqnarray}
  where 
 \begin{equation}
     D_{x,y} = \left( \begin{array}{cccc} |x|^2 & xy^* & -|x|^2 & - x y^* \\ yx^* & |y|^2 & - yx^* & - |y|^2 \\ - |x|^2 & -xy^* & |x|^2 & xy^* \\ - yx^* & - |y|^2 & yx^* & |y|^2 \end{array} \right).
 \end{equation}
   Now,  our idea is to show that the determinant of
$\Phi_{\theta,\lambda,x,y} (|\psi\rangle\langle\psi|)$  is negative for appropriately choosen $|\psi\rangle$, if $\lambda>0$  and $x,y$ are arbitrary.
  Hence we  claim that no rank-1 projector (hence no positive operators) can be subtracted from the EW $W_{II}(\theta)$ in this class and this witness is optimal without having the spanning property for $\theta \in (0,\pi)$.
  
For this purpose, let us consider the following  vector 
\begin{equation}
\label{vec1}
|\psi_{\lambda,k}\rangle=\left(
\begin{array}{c}
 0 \\
 1 \\
 \sqrt{2k \sin{\frac{\theta}{2}} \lambda} \\
 1+k \cos\frac{\theta}{2} \lambda \\
\end{array}
\right),
\end{equation}
 with the correction $\lambda$ and arbitrary small parameter $k>0$.  The action of the positive map on the projector onto this vector is
\begin{eqnarray}
\label{dekhi2}
\Phi_\theta(|\psi_{\lambda,k}\rangle \langle \psi_{\lambda,k}|)-\lambda D_{x,y} \circ |\psi_{\lambda,k}\rangle \langle \psi_{\lambda,k}| . \nonumber\\
\end{eqnarray}
   Now, we  show that the determinant of the above operator is negative.  In order to do so, we consider its determinant as a series of powers of $\lambda$. The leading power is 3 and the corresponding coefficient reads
   \begin{equation}
   8 k^2 \big(k \sin\frac{\theta}{2} \cos^2\frac{\theta}{2}-|x|^2 \sin^2\frac{\theta}{2}\big).
   \end{equation}
Now it is clear, that for any value of $\theta$  and for any non-zero value of $x$ one can choose $k$  small enough to make the leading coefficient negative, hence the determinant is negative for small enough $\lambda$.
Clearly here we cannot exclude the correction for  $x=0,~y=1$ in this way. However, in this case the correction can be excluded by considering the same vector (\ref{vec1}), but with permuted entries. This ends the proof.

\end{document}